\documentclass[preprint,showpacs,preprintnumbers,amsmath,amssymb]{revtex4}

\usepackage{graphicx}
\usepackage{dcolumn}
\usepackage{bm}
\usepackage{color}
\usepackage{hyperref}
\usepackage{amssymb}
\usepackage{amsmath,empheq}
\usepackage{slashed}

\usepackage{tikz}
\usepackage{xcolor}
\usepackage[normalem]{ulem}

\makeatletter
\def\l@subsection#1#2{}
\def\l@subsubsection#1#2{}
\makeatother
\begin{document}

\preprint{OU-HET-1007}

\title{TKNN formula for general Hamiltonian}

\author{Hidenori Fukaya}
\email{hfukaya@phys.sci.osaka-u.ac.jp}
 \affiliation{Department of Physics, Osaka University,
 Toyonaka, Osaka 560-0043, Japan.}
\author{Tetsuya Onogi}
\email{onogi@phys.sci.osaka-u.ac.jp}
\affiliation{Department of Physics, Osaka University,
 Toyonaka, Osaka 560-0043, Japan.}
 \author{Satoshi Yamaguchi}
\email{yamaguch@het.phys.sci.osaka-u.ac.jp}
\affiliation{Department of Physics, Osaka University,
 Toyonaka, Osaka 560-0043, Japan.}
 \author{Xi Wu}
\email{wuxi5949@gmail.com}
\affiliation{Physics Department, Ariel University, Ariel 40700, Israel.}

\begin{abstract}
Topological insulators in odd dimensions are characterized by topological numbers. 
We prove the well-known relation between the topological number given by the Chern character of the Berry curvature and the Chern-Simons level of the low energy effective action for a general class of Hamiltonians bilinear in the fermion with general U(1) gauge interactions including non-minimal couplings by an explicit calculation. A series of Ward-Takahashi identities are crucial to relate the Chern-Simons level to a winding number, which could then be directly reduced to Chern character of Berry curvature by carrying out the integral over the temporal momenta. 
\end{abstract}

\maketitle


\section{Introduction}

Topological insulators in $D=2n+1$ dimensions are characterized by topological numbers. One characterization is given by the Chern character of the Berry connection from the eigenfunctions of the Hamiltonian in the valence band~\cite{Thouless:1982aa, Niu:1985aa}, the other characterization is given by the level of the Chern-Simons action which appears in the effective action after integrating out the fermion coupled to a smooth external U(1) gauge field, i.e., photon~\cite{Ishikawa:1983ad,Ishikawa:1984zv,Ishikawa:1986wx,Ishikawa:1987zi,So:1984nf,So:1985wv,Golterman:1992ub}.
 These two characterization are known to be equivalent because they both arise from the current correlation functions and there are explicit proofs for various cases.

For example, the famous TKNN number \cite{Thouless:1982aa} describes integer Hall conductivity in spatial two-dimensional systems and it is valid for all kinds of band structures neglecting interactions between electrons. Meanwhile, 
motivated by the discovery of domain-wall fermion \cite{Kaplan:1992bt}  (see also subsequent papers to study the anomaly inflow 
\cite{Jansen:1992yj, Jansen:1992tw, Kaplan:1995pe})
in their paper, Golterman Jansen and Kaplan (GJK) also gave an expression of conductivity of Chern-Simons current 
for a wide class of  fermion propagator on the lattice including Wilson fermion 
with odd-dimensional Euclidean spacetime and found out that the topological number is given by the homotopy class for the map $T^D \rightarrow S^D$\cite{Golterman:1992ub}. 
They also showed that the topological number one finds correlated perfectly with the number of chiral edge states
. The relation is well-known to many people, and in the continuum theory or in some special models there exist studies for the relations \cite{Witten3lects,Zubko:2016298}. 

In Ref.~\cite{Qi:2008ew} a proof is given for a large class of models for general odd dimensions, where they consider the most general lattice action for arbitrary free kinetic term on the lattice which is then coupled to U(1) gauge field in a minimal way, i.e. with the gauge interaction in the form of 
\begin{eqnarray}
H(A) = \sum_{m,n} \psi^\dagger_m h_{mn} e^{iA_{mn}} \psi_n + \sum_m \psi_m^\dagger \psi_m,
\end{eqnarray}
where $m,n$ are the lattice sites $h_{mn}$ are the hopping parameters and $A_{mn}$ is the line integral of the gauge field along the straight line connecting the sites $m$ and $n$.  The advantage of this class of Hamiltonian is that the contact interactions such as fermion-fermion-multi-photon vertices do not contribute to the final expression so that only a set of Feynman diagram which appear  also in the continuum theory gives non-vanishing contributions. Of course, this type of gauge interaction is physically motivated since it is based on the famous method of `Peierls substitution'~\cite{Peierls:1933}. However,  in more general situation, the gauge interaction may not always be described by such a single straight Wilson-line. It could be a linear combination of various Wilson-lines of arbitrary path, which can give non-minimal coupling. In such cases, one has to include the contribution of contact interaction vertices. 

In this paper, we 
study the 
equivalence of topological number from TKNN formula 
and that from the Chern-Simons coupling for the most general lattice fermion Hamiltonian coupled to U(1) gauge field which is bilinear in fermion. The new features of our study is that the Hamiltonian is  general enough to include arbitrary non-minimal gauge interactions which has not been considered in the previous studies\cite{Ishikawa:1983ad,Ishikawa:1984zv,Ishikawa:1986wx,Ishikawa:1987zi,So:1984nf,So:1985wv,Golterman:1992ub,Witten3lects,Zubko:2016298,Qi:2008ew} 
. 
We give an explicit proof of the equivalence of the two topological numbers for gapped fermion systems with Hamiltonian on the lattice given by the bilinear form of the fermion coupled to external U(1) fermions. We also discuss  how the relation can give the Chern character in higher dimensional case.

The organization of this paper is as follows. In the next section we rewrite the  level 
of the Chern-Simons effective action for the gapped fermion system coupled to a U(1) gauge field 
using the 
Feynman rule
and relate it to the  winding number of a map from $T^D$ to the fermion propagator space in $D=2+1$ and $D=4+1$ dimensions. In Section 3, we show the equivalence of the winding number to the Chern number for the Berry curvature.  Section 4 is devoted to summary and conclusion. 

\section{Gapped fermion system on the lattice}

\subsection{General gapped fermion system}
We consider a gapped fermion system on a lattice (or condensed matter systems on a translational invariant crystal) with the following action in Euclidean space 
in $D=2n+1$ dimensions. (Note that the time is continuous but the space is discrete as in the condensed matter systems.)
\begin{eqnarray}
S_{\rm E} = \int dt \sum_{\vec{r}}\psi^\dagger(t,\vec{r}) 
\left[\frac{\partial}{\partial t}  + i A_0 + H(\vec{A})\right] \psi(t,\vec{r}),
\end{eqnarray}
where $\vec{r}$ runs over the $2n$ dimensional spatial lattice points. 
We will set $x^0=t$ in the followings. 
The Hamiltonian $H(\vec{A})$ is given by a summation over all the possible hoppings on the lattice 
which include
gauge interactions with a smooth external U(1) gauge field $A_\mu=(A_0,\vec{A})$. The fermion fields $\psi^\dagger(t,\vec{r})$ and $\psi(t,\vec{r})$ give creation and annihilation operators of  fermions after quantization. We assume that when the gauge field is turned off, the Hamiltonian is translational invariant so that it can allow band structures. We also assume that there are $N_v$ bands and $N_c$ bands below and above the fermi level, respectively.  Therefore the fermion fields have $N_v+N_c$ components. 

\subsection{Effective gauge action}
Since the fermion system is gapped with a gap size 
$\Delta >0$
, the effective gauge action 
obtained by
integrating out fermions can be expanded in terms of gauge invariant local actions as
\begin{eqnarray}
S_{\rm eff}= \sum_{k} a_k S_k(A).
\end{eqnarray}
Here, $S_{\rm eff}(A)$ is defined as 
\begin{eqnarray}
e^{S_{\rm eff}} = \int \mathcal{D}\psi\mathcal{D}\psi^\dagger e^{-S_{\rm E}},
\end{eqnarray}
and $S_k(A)$ are the gauge invariant actions given by the local Lagrangian $L_k(A)$ and $a_k$ are
the coefficients. By dimensional analysis, if the Lagragian $L_k(A)$ has a mass dimension $d_k$
the coefficient $a_k$ is suppressed by the $d_k-(2n+1)$-powers in $\frac{1}{\Delta}$ or lattice spacing $a$. Many of the Lagrangians are given in terms of gauge invariant field $F_{\mu\nu}= \partial_\mu A_\nu- \partial_\nu A_\mu$, (e.g. 
$S^{F^2}_{\mu\nu,\rho\sigma} \equiv \int d^{2n+1} r ~F_{\mu\nu} F_{\rho\sigma}$ with a coefficent  $a^{F^2}_{\mu\nu, \rho\sigma}$). Since we do not have the Lorentz-invariance on the lattice,  the structure of the coefficients $a_k$ in the effective action heavily depend the geometry of the lattice. 

However, there is a very special parity-violating term called Chern-Simons action $S_{\rm cs}(A)$ given by
\begin{eqnarray}
S_{\rm cs}(A) = \int d^{2n+1} x ~ \epsilon_{\alpha_0 \beta_1\alpha_1 \cdots \beta_{n}\alpha_{n}}
A_{\alpha_0} \partial_{\beta_1} A_{\alpha_1} \cdots 
\partial_{\beta_{n}} A_{\alpha_{n}} .
\end{eqnarray}
This action is topological and always takes this form no matter what the geometry of the lattice is.
Topological information of the fermion system is contained in the effective action through the coefficient $c_{\rm cs}$ as
\begin{eqnarray}
S_{\rm eff}(A) =  i c_{\rm cs} S_{\rm cs} + \mbox{\rm  other gauge invariant terms} .
\end{eqnarray}
Here the gauge invariance of the action requires that the coefficient is quantized as 
(See \cite{Witten3lects} for example.)
\begin{eqnarray}
c_{\rm cs} = \frac{k}{(2\pi)^n (n+1)!}  &, ~~& k\in \mathbb{Z} .
\label{eq:ccs_k}
\end{eqnarray}
Since the Chern-Simons action is of the lowest dimension in the parity-violating sector, the coefficient $c_{\rm cs}$ can be obtained by the following  quantity
\begin{eqnarray}
c_{\rm cs}
&=&
\frac{(-i)^{n+1} \epsilon_{\alpha_0\beta_1 \alpha_1\cdots \beta_n \alpha_n}}{(n+1)!(2n+1)!}
\left(\frac{\partial}{\partial (q_1)_{\beta_1}}\right)
\cdots
\left(\frac{\partial}{\partial (q_n)_{\beta_n}}\right)
\nonumber\\
&&
\times
\left.
\prod_{i=1}^n\int d^D x_i
e^{iq_i x_i}
\frac{\delta^{n+1} S_{\rm eff}(A)}
{\delta A_{\alpha_0}(x_0) \delta  A_{\alpha_1}(x_1)
\cdots \delta A_{\alpha_n}(x_n)}
\right|_{q_i=0} .
\label{eq:Ccs_Seff}
\end{eqnarray}

In the continuum theory or in simple lattice fermions such as Wilson fermion, this can be rewritten in terms of fermion path-integral as
\begin{eqnarray}
c_{\rm cs}
&=&
-\frac{(-i)^{n+1} \epsilon_{\alpha_0\beta_1\alpha_1 \cdots \beta_n \alpha_n}}{(n+1)!(2n+1)!}
~ n! ~ \left(\frac{\partial}{\partial (q_1)_{\beta_1}}\right)
\cdots
\left(\frac{\partial}{\partial (q_n)_{\beta_n}}\right)
\times
\int\frac{dp_0}{2\pi}
\int_{\rm BZ}
\frac{d^{2n}p}{(2\pi)^{2n}}
\nonumber\\
&&
\left. 
\mbox{Tr}
\left[
S_F(p+q_{n+1})\Gamma^{(1)}[q_{n+1}, \alpha_0;p]S_F(p)\Gamma^{(1)}[q_1,\alpha_1;p-q_1] S_F(p-q_1)
\cdots\Gamma^{(1)}(q_n,\alpha_n;p)
\right]
\right|_{q_i=0} .
\nonumber\\
\end{eqnarray}
Here, 
BZ stands for the Brillioune zone, 
 $q_{n+1}\equiv q_1+\cdots+q_n$,
$S_F(p)$ is the Fourier transform of the free fermion propagator $\displaystyle{\frac{1}{\frac{\partial}{\partial x_0} +H(\vec{A}=0)}}$ with momentum $p$, and $\Gamma^{(1)}[q,\alpha;p]$ is  the fermion-fermion-photon vertex with in-coming fermion momentum $p$ and in-coming photon momentum $q$.
However, the situation is not simple in general, since there are also contributions from contact interactions such as fermion-fermion-multi-photon vertices, which can naturally arise from non-minimal gauge couplings or generic lattice artifacts.
In the following, we will explicitly show that these contributions automatically cancel against the contributions from the momentum derivative of vertex functions due to the Ward-Takahashi identities.

In the next subsection, we formulate how to evaluate Eq.(\ref{eq:Ccs_Seff}) using fermion propagators and vertex functions for general Hamiltonian system.


\subsection{Fermion propagator representation of Eq.(\ref{eq:Ccs_Seff})}

The effective action can be given by the log of the fermion determinant as
\begin{eqnarray}
S_{\rm eff}(A) = \mbox{Tr}\left[\ln\left(D_0 + H(A)\right)\right],
\label{eq:S=lnDet}
\end{eqnarray}
where $D_0= \frac{\partial}{\partial x^0}+i A_0$.
Splitting the kinetic operator $D_0 + H(A)$ into free part and interaction part as
\begin{eqnarray}
D_0 + H(A) = \frac{\partial}{\partial x_0} + H_0 - \Gamma(A), 
\label{eq:split_kin}
\end{eqnarray}
where $H_0$ is the free fermion part defined as $H_0\equiv \left. H(A)\right|_{A=0}$ and $\Gamma(A)$ is the interaction part defined as 
$\Gamma(A) \equiv - iA_0 - H(A) + H_0$.
Plugging Eq.(\ref{eq:split_kin}) into Eq.(\ref{eq:S=lnDet}) we obtain
\begin{eqnarray}
S_{\rm eff}(A) -\mbox{\rm const.}&=& - \sum_{n=1}^\infty \frac{1}{n}\mbox{Tr}\left[\left(\frac{1}{\frac{\partial}{\partial x_0} + H_0}\Gamma(A)\right)^n\right] 
\end{eqnarray}
Simple algebra shows that the following 
equation holds:
\begin{eqnarray}
&&
\left. \epsilon_{\alpha_0\beta_1\alpha_1}\frac{\delta^2 S_{\rm eff}}{\delta A_{\alpha_0}(x_0) 
\delta A_{\alpha_1}(x_1)}
 \right|_{A=0}
\nonumber\\
&=& 
-\epsilon_{\alpha_0\beta_1\alpha_1}
\left\{
\left.\mbox{Tr}\left[\frac{1}{\frac{\partial}{\partial x_0} + H_0} \cdot\frac{\delta^2\Gamma(A)}{\delta A_{\alpha_0}(x_0)\delta A_{\alpha_1}(x_1)}\right]\right|_{A=0}
\right.
\nonumber\\
&&
\left. \hspace{1.5cm} +
\left.\mbox{Tr}\left[\frac{1}{\frac{\partial}{\partial x_0} + H_0} \cdot\frac{\delta\Gamma(A)}{\delta A_{\alpha_0}(x_0)}
\cdot\frac{1}{\frac{\partial}{\partial x_0} + H_0} \cdot\frac{\delta\Gamma(A)}{\delta A_{\alpha_1}(x_1)}\right]\right|_{A=0}
\right\},
\label{eq:delS_AA}
\end{eqnarray}
From Eq.(\ref{eq:delS_AA}), we find that the Chern-Simons 
coupling
for $D=2+1$ dimension is given by
\begin{eqnarray}
c_{\rm cs} 
&=& 
-\frac{(-i)^2\epsilon_{\alpha_0\beta_1\alpha_1}}{2!3!}\int \frac{d^3p}{(2\pi)^3}
\left(\frac{\partial}{\partial q_1}\right)_{\beta_1}
\nonumber\\
&&
\left\{
\mbox{Tr}\left[S_F(p) 
\Gamma^{(2)}[-q_1,\alpha_0;q_1,\alpha_1;p]\right]\right.
\nonumber\\
&&
 +
\left.
\left.\mbox{Tr}\left[S_F(p-q_1) 
\Gamma^{(1)}[-q_1,\alpha_0;p]
 S_F(p) 
  \Gamma^{(1)}[q_1,\alpha_1;p-q_1] \right]
\right\}\right|_{q_1=0},
\end{eqnarray}
where $S_F(p)$ is the fermion propagator $\frac{1}{ip_0 +H_0(\vec{p})}$ and $\Gamma^{(1)}[q_1,\alpha_1;p]$ and $\Gamma^{(2)}[q_1,\alpha_1;q_2,\alpha_2;p]$ 
are fermion-fermion-photon and fermion-fermion-photon-photon vertices  with in-coming fermion momentum $p$ and in-coming photon momenta $q_i ~(i=1,2)$ with Lorentz index 
$\alpha_i ~(i=1,2)$
\begin{eqnarray}
\Gamma^{(1)}[q_1,\alpha_1;p]
&=& \int d^{2n+1} x_1 ~e^{iq_1\cdot x_1}\int d^{2n+1} y ~ e^{ip \cdot y}
\left. \frac{\delta \Gamma[A](x,y)}{\delta A_{\alpha_1}(x_1)} \right|_{A=0, x=0},
\\
\Gamma^{(2)}[q_1,\alpha_1,q_2;\alpha_2;p]
&=& \prod_{i=1}^2\left(\int d^{2n+1} x_i~ e^{iq_i\cdot x_i}\right)\int d^{2n+1} y ~ e^{ip \cdot y}
\left. \frac{\delta^2 \Gamma[A](x,y)}{\delta A_{\alpha_1}(x_1)\delta A_{\alpha_2}(x_2)} \right|_{A=0, x=0}.
\nonumber\\
\end{eqnarray}
Note that the contributions with multi-photon vertices vanishes for the class of Hamiltonians with gauge interactions given by a single straight Wilson-line because the multi-photon vertices are symmetric under the interchange of Lorentz indices of photons. When contracted with the antisymmetric tensor, such contributions vanish. However, in general Hamiltonian we must consider these contributions.

From a similar calculation,
we find that the Chern-Simons level for $D=4+1$ dimension is given by
\begin{eqnarray}
&&c_{\rm cs} 
= 
-\frac{(-i)^3\epsilon_{\alpha_0\beta_1\alpha_1\beta_2\alpha_2}}{3!5!}\int \frac{d^5p}{(2\pi)^5}
\left(\frac{\partial}{\partial q_1}\right)_{\beta_1}
\left(\frac{\partial}{\partial q_2}\right)_{\beta_2}
\nonumber\\
&\left\{ \right.&
\mbox{Tr}\left[S_F(p) \cdot \Gamma^{(3)}[-(q_1+q_2),\alpha_0;q_1,\alpha_1;q_2,\alpha_2;p]\right]
\nonumber\\
&+&
2\mbox{Tr}\left[S_F(p-q_2) 
\Gamma^{(2)}[-(q_1+q_2),\alpha_0;q_1,\alpha_1;p]
S_F(p) 
 \Gamma^{(1)}[q_2,\alpha_2;p-q_2] \right]
\nonumber\\
&+&
\left.
\mbox{Tr}\left[S_F(p+q_1+q_2)
  \Gamma^{(2)}[q_1,\alpha_1;q_2,\alpha_2;p]
S_F(p)
  \Gamma^{(1)}[-(q_1+q_2),\alpha_0;p+q_1+q_2] \right]
\right\},
\nonumber\\
&+&
\left.
\left.
2\mbox{Tr}\left[S_F(p+q_1) 
\Gamma^{(1)}[q_1,\alpha_1;p]
S_F(p)  
\Gamma^{(1)}[q_2,\alpha_2;p-q_2] \
S_F(p-q_2)  
\Gamma^{(1)}[-(q_1+q_2),\alpha_0;p+q_1] \right]
\right\}\right|_{q_1=q_2=0},
\nonumber\\
\end{eqnarray}
where $\Gamma^{(3)}[q_1,\alpha_1;q_2,\alpha_2;q_3,\alpha_3;p]$ 
is the fermion-fermion-photon-photon-photon vertex with in-coming fermion momentum $p$ and in-coming photon momenta $q_i ~(i=1,2,3)$ with Lorentz index 
$\alpha_i ~(i=1,2,3)$
, which is given by
\begin{eqnarray}
\Gamma^{(3)}[q_1,\alpha_1,q_2;\alpha_2.q_3,\alpha_3;p]
&=& \prod_{i=1}^3\left(\int d^{2n+1} x_i~ e^{iq_i\cdot x_i}\right)\int d^{2n+1} y ~ e^{ip \cdot y}
\left. \frac{\delta^3 \Gamma[A](x,y)}{\delta A_{\alpha_1}(x_1)\delta A_{\alpha_2(x_2)}\delta A_{\alpha_3(x_3)}} \right|_{A=0, x=0}.
\nonumber\\
\end{eqnarray}


\subsection{The case for $D=2+1$ dimension~($n=1$)}
In $D=2+1$ dimension, i.e. $n=1$ case, the Chern-Simons 
coupling
$c_{\rm cs}$ has contributions from 
the loop involving two fermion-fermion-photon vertices and the loop involving a single fermion-fermion-photon-photon vertex (contact interaction) as
\begin{eqnarray}
c_{\rm cs} 
&=& 
-\frac{(-i)^2}{2!3!}\int \frac{d^3p}{(2\pi)^3} X,
\end{eqnarray}
 where  $X$ is given as
\begin{eqnarray}
X
&=& 
\epsilon_{\alpha_0\beta_1\alpha_1}
\left(\frac{\partial}{\partial q_1}\right)_{\beta_1}\left\{
\mbox{Tr}\left[S_F(p) 
\Gamma^{(2)}[-q_1,\alpha_0;q_1,\alpha_1;p]\right]\right.
\nonumber\\
&&
 +
\left.\mbox{Tr}\left[S_F(p-q_1) 
\Gamma^{(1)}[-q_1,\alpha_0;p]
 S_F(p) 
  \Gamma^{(1)}[q_1,\alpha_1;p-q_1] \right]
\right\}.
\end{eqnarray}
The first term on the right hand side is the one-loop contribution with contact interaction and the second term is the usual one-loop contribution with simple fermion-fermion-photon vertices.

Carrying out the momentum derivative with $q$, 
\begin{eqnarray}
X
&=&
\epsilon_{\alpha_0\beta_1 \alpha_1}
\times
\left\{ 
\mbox{Tr}
\left.
\left[
S_F(p)
\left(
-\frac{\partial \Gamma^{(2)}[q,\alpha_0;0,\alpha_1;p]}{\partial q_{\beta_1}}
+\frac{\partial \Gamma^{(2)}[0,\alpha_0;q, \alpha_1;p]}{\partial q_{\beta_1}}
\right)
\right]
\right|_{q=0} 
\right.
\nonumber\\
&&
+
\mbox{Tr}
\left.
\left[
S_F(p)
\left(\frac{\partial \Gamma^{(1)}[q,\alpha_1;p]}{\partial q_{\beta_1}} 
-\ \frac{\partial \Gamma^{(1)}[0,\alpha_1;p]}{\partial p_{\beta_1}}
\right)
S_F(p)
\Gamma^{(1)}(0,\alpha_0;p)
\right]
\right|_{q=0} 
\nonumber\\
&&
\left.
-\mbox{Tr}
\left[
S_F(p)
 \Gamma^{(1)}[0,\alpha_1;p]
S_F(p)
\frac{\partial \Gamma^{(1)}(q,\alpha_0;p)}{\partial q_{\beta_1}}
\right]
\right|_{q=0} 
\nonumber\\
&&
\left.\left.
-\mbox{Tr}
\left[
S_F(p)
 \Gamma^{(1)}[0,\alpha_1;p]
\frac{\partial S_F(p)}{\partial p_{\beta_1}}
\Gamma^{(1)}(0,\alpha_0;p)
\right]
\right|_{q=0} 
\right\}
\textcolor{red}{.}
\nonumber\\
\end{eqnarray}
In 
Appendix~ \ref{App_WT}
, we derive 
the Ward-Takahashi identities as follows:
\begin{eqnarray}
\Gamma^{(1)}[0,\alpha;p] = \textcolor{blue}{-}\frac{\partial S^{-1}_F(p)}{\partial p_\alpha},
\label{eq:1st_WT}
\end{eqnarray}
\begin{eqnarray}
\left.\frac{\partial^2 \Gamma^{(1)}[k,\mu;p]}{\partial k_\nu \partial p_\lambda}\right|_{k=0}
=\left. \frac{\partial \Gamma^{(2)}[k,\mu;0,\lambda;p]}{\partial k_\nu}\right|_{k=0}
=\left. \frac{\partial \Gamma^{(2)}[0,\lambda;l,\mu;p]}{\partial l_\nu}\right|_{l=0} .
\label{eq:2nd_WT}
\end{eqnarray}
Using these identities, we obtain
\begin{eqnarray}
X
&=&
\epsilon_{\alpha_0\beta_1 \alpha_1}
\times
\left\{ 
2\mbox{Tr}
\left.
\left[
S_F(p)
\left(
\frac{\partial^2 \Gamma^{(1)}[q, \alpha_1;p]}{\partial q_{\beta_1} \partial p_{\alpha_0}}
\right)
\right]
\right|_{q=0} 
\right.
+2
\mbox{Tr}
\left.
\left[
\frac{\partial S_F(p)}{\partial p_{\alpha_0}}
\frac{\partial \Gamma^{(1)}[q, \alpha_1;p]}{\partial q_{\beta_1} }
\right]
\right|_{q=0} 
\nonumber\\
&&
\left.\left.
+\mbox{Tr}
\left[
S_F(p)
\frac{\partial S^{-1}_F(p)}{\partial p_{\alpha_1}}
S_F(p)
\frac{\partial S^{-1}_F(p)}{\partial p_{\beta_1}}
S_F(p)
\frac{\partial S^{-1}_F(p)}{\partial p_{\alpha_0}}
\right]
\right|_{q=0} \right\}
\textcolor{red}{.}
\end{eqnarray}
The first and the second terms on the right hand side can be combined to give a total divergence which  vanishes when we integrate over the momentum. Therefore, one finds that the Chern-Simons 
coupling
is given by the winding number as
\begin{eqnarray}
c_{\rm cs}
&=&
\frac{(-i)^2 \epsilon_{\alpha_0\beta_1 \alpha_1}}{2!3!}
\int\frac{dp_0}{2\pi}
\int_{\rm BZ}
\frac{d^{2}p}{(2\pi)^{2}}
\mbox{Tr}
\left[
S_F(p)
\frac{\partial S_F^{-1}(p)}{\partial p_{\alpha_0}}
S_F(p)\frac{\partial S^{-1}_F(p)}{\partial p_{\beta_1}}
S_F(p)\frac{\partial S_F^{-1}(p)}{\partial p_{\alpha_1}}
\right].
\end{eqnarray}

\subsection{The case for $D=4+1$ dimension~($n=2$)}
The Chern-Simons 
coupling
 can be
  written
  as
\begin{eqnarray}
&&c_{\rm cs} 
= 
-\frac{(-i)^3}{3!5!}\int \frac{d^5p}{(2\pi)^5}\left[X_1+X_2+X_3+X_4\right],
\end{eqnarray}
where $X_1, X_2, X_3, X_4$ are defined as follows
\begin{eqnarray}
X_1&=&
\epsilon_{\alpha_0\beta_1\alpha_1\beta_2\alpha_2}
\left(\frac{\partial}{\partial q_1}\right)_{\beta_1}
\left(\frac{\partial}{\partial q_2}\right)_{\beta_2}
\left.
\left\{\mbox{Tr}\left[S_F(p) \cdot \Gamma^{(3)}[-(q_1+q_2),\alpha_0;q_1,\alpha_1;q_2,\alpha_2;p]\right]\right\}
\right|_{q_1=q_2=0}
\nonumber\\
X_2&=&
\epsilon_{\alpha_0\beta_1\alpha_1\beta_2\alpha_2}
\left(\frac{\partial}{\partial q_1}\right)_{\beta_1}
\left(\frac{\partial}{\partial q_2}\right)_{\beta_2}
\nonumber
\\
&&
\left.
\left\{2\mbox{Tr}\left[S_F(p-q_2) 
\Gamma^{(2)}[-(q_1+q_2),\alpha_0;q_1,\alpha_1;p]
S_F(p) 
 \Gamma^{(1)}[q_2,\alpha_2;p-q_2] \right]
 \right\}
\right|_{q_1=q_2=0},
\nonumber\\
X_3&=&
\epsilon_{\alpha_0\beta_1\alpha_1\beta_2\alpha_2}
\left(\frac{\partial}{\partial q_1}\right)_{\beta_1}
\left(\frac{\partial}{\partial q_2}\right)_{\beta_2}
\nonumber
\\
&&
\left.
\left\{
\mbox{Tr}\left[S_F(p+q_1+q_2)
  \Gamma^{(2)}[q_1,\alpha_1;q_2,\alpha_2;p]
S_F(p)
  \Gamma^{(1)}[-(q_1+q_2),\alpha_0;p+q_1+q_2] \right]
\right\}
\right|_{q_1=q_2=0},
\nonumber\\
X_4&=&
\epsilon_{\alpha_0\beta_1\alpha_1\beta_2\alpha_2}
\left(\frac{\partial}{\partial q_1}\right)_{\beta_1}
\left(\frac{\partial}{\partial q_2}\right)_{\beta_2}
\nonumber
\\
&&
\left.
\left\{
2\mbox{Tr}\left[S_F(p+q_1) 
\Gamma^{(1)}[q_1,\alpha_1;p]
S_F(p)  
\Gamma^{(1)}[q_2,\alpha_2;p-q_2] \
S_F(p-q_2)  
\right.
\right.
\right.
\nonumber\\
&&
\left.
\left.
\left.
\hspace{1cm}\Gamma^{(1)}[-(q_1+q_2),\alpha_0;p+q_1] \right]
\right\}\right|_{q_1=q_2=0}.
\nonumber\\
\end{eqnarray}

Using eqs.(\ref{eq:1st_WT}, \ref{eq:2nd_WT}) as in the case of $D=2+1$ dimensions.
as well as 
the following Ward-Takahashi 
 identities given in Appendix\ref{App_WT} , 
\begin{eqnarray}
\left.\frac{\partial^2 \Gamma^{(3)}[q,\mu; r,\nu;0,\lambda;p]}{\partial q_{\alpha} \partial r_{\beta}}\right|_{q=r=0}
= \left.\frac{\partial^3\Gamma^{(2)}[q,\mu; r,\nu;p]}{\partial q_{\alpha} \partial r_{\beta} \partial p_\lambda}\right|_{q=r=0}.
\label{eq:WT_G3}
\end{eqnarray}
we obtain
\begin{eqnarray}
&&X_1+X_2+X_3+X_4
\nonumber\\
&=&\epsilon_{\alpha_0\beta_1\alpha_1\beta_2\alpha_2} 
\nonumber\\
&&
\left[
\frac{\partial}{\partial p_{\alpha_0}}\left\{
3\mbox{Tr}
\left[
\left. 
\frac{
\Gamma^{(2)}[q,\alpha_1;r,\alpha_2;p]
}{\partial q_{\beta_1} r_{\beta_2} }
\right|_{q=r=0}
 S_F(p)
\right]
\right.
\right.
\nonumber\\
&&
\hspace{1.2cm}
+
3\mbox{Tr}
\left[
\left. 
\frac{\Gamma^{(1)}[q,\alpha_1;p]}{\partial q_{\beta_1} }
\right|_{q=0}
 S_F(p)
 \left. 
\frac{
\Gamma^{(1)}[q^\prime,\alpha_2;p]
}{\partial q^\prime_{\beta_2}  }
\right|_{q^\prime=0}
S_F(p)
\right]
\nonumber\\
&&
\left.
\hspace{1.2cm}
+ 8 \mbox{Tr}
\left[
\left. 
\frac{\Gamma^{(1)}[q,\alpha_1;p]}{\partial q_{\beta_1} }
\right|_{q=r=0}
 S_F(p)
\frac{\partial S^{-1}_F(p)}{\partial p_{\alpha_2}}
\frac{\partial S_F(p)}{\partial p_{\beta_2}}
\right]
\right\}
\nonumber\\
&&
+2
\mbox{Tr}
\left[
\left. 
S_F(p)
\frac{\partial S^{-1}_F(p)}{\partial p_{\alpha_0}}
 S_F(p)
\frac{\partial S^{-1}_F(p)}{\partial p_{\beta_1}}
 S_F(p)
\frac{\partial S^{-1}_F(p)}{\partial p_{\alpha_1}}
 S_F(p)
\frac{\partial S^{-1}_F(p)}{\partial p_{\beta_2}}
 S_F(p)
\frac{\partial S^{-1}_F(p)}{\partial p_{\alpha_2}}
\right]
\right].
\nonumber\\
&&
\end{eqnarray}
The total divergence term will vanish after integrating over the spatial momenta due to the periodicity 
in BZ.

Thus, we finally get 
\begin{eqnarray}
c_{\rm cs} 
&=& 
-
\frac{(-i)^3\cdot 2}{3!5!}\int \frac{d^5p}{(2\pi)^5}
\epsilon_{\alpha_0\beta_1\alpha_1\beta_2\alpha_2} 
\nonumber\\
&&
\mbox{Tr}
\left[
S_F(p)
\frac{\partial S^{-1}_F(p)}{\partial p_{\alpha_0}}
 S_F(p)
\frac{\partial S^{-1}_F(p)}{\partial p_{\beta_1}}
 S_F(p)
\frac{\partial S^{-1}_F(p)}{\partial p_{\alpha_1}}
 S_F(p)
\frac{\partial S^{-1}_F(p)}{\partial p_{\beta_2}}
 S_F(p)
\frac{\partial S^{-1}_F(p)}{\partial p_{\alpha_2}}
\right].
\end{eqnarray}

Therefore, Chern-Simons coupling is given by the winding number with fermion propagator also for $D=4+1$ case.

We expect that the relation of Chern-Simons coupling and the winding number for general Hamiltonian including non-minimal coupling 
holds for arbitrary odd dimensions ($D=2n+1$). This will be left for future studies.

\section{Equivalence of  winding number and chern number}
In this section, we show the equivalence of the Chern-Simons coupling given by the winding number expression and the Chern character given by the Berry connection 
for the energy eigenstates in the valence bands. 

The proof of this part is already given in Ref.~\cite{Qi:2008ew}, but since the proof is simple, we give it here for completeness.
We give the calculation for arbitrary odd ($D=2n+1$) dimensions, even though we have shown that the Chern-Simons coupling $c_{\rm cs}$  
can be written by the winding number using $S_F$ only for $D=2+1$ and $D=4+1$ dimensions.

In order to simplify the notation, hereafter we abbreviate the derivative with respect to the momentum $p_\mu$ 
as $\partial_\mu \equiv \frac{\partial}{\partial p_\mu}$.
\subsection{Winding number in $D=2n+1$ dimension}
The result of the previous section for $D=2+1$ and $D=4+1$ can be unified to the following results:
In the expression using the fermion propagator $S(p)$, the Chern-Simons coupling $c_{\rm cs}$ in $D=2n+1$ dimensions is given as
\begin{eqnarray}
	c_{\rm cs}&=&-\frac{n!\cdot (2n+1) (-i)^{n+1}\epsilon^{i_1 i_2\cdots i_{2n}}}{(n+1)!(2n+1)!}\int \frac{d^{2n}p}{(2\pi)^{2n}}\int \frac{dp^0}{2\pi}
\text{Tr}\left[S_F(\partial_0S_F^{-1})\prod_{k=1}^{2n}\left(S_F(\partial_{i_k}S_F^{-1})\right)\right]
\label{eq:winding_2n+1}
\nonumber\\
	&=&-\frac{n!\cdot (2n+1) (-i)^{n+1}\epsilon^{i_1 i_2\cdots i_{2n}}}{(n+1)!(2n+1)!}\int \frac{d^{2n}p}{(2\pi)^{2n}}\int \frac{dp^0}{2\pi}
\text{Tr}\left[\frac{1}{ip^0+H} ~i~\prod_{k=1}^{2n}\left(\frac{1}{ip^0+H}(\partial_{i_k}H)\right)\right]
\nonumber\\
\end{eqnarray}
Next we insert a complete set $\displaystyle{\sum_{\alpha}|\alpha\rangle\langle \alpha |}$, where $\alpha$ is the label of energy.
Then we have:
\begin{eqnarray}
	c_{\rm cs}
	&=&\frac{n! (-i)^{n+2}}{(n+1)!(2n)!}\int \frac{d^{2n}p}{(2\pi)^{2n}} J .
\label{eq:ccs_J2n+1}
\end{eqnarray}
Here $J$ is defined as 
\begin{eqnarray}
J = \sum_{\alpha_1,\cdots, \alpha_{2n}} \epsilon^{i_1 i_2 \cdots i_{2n}}
\int \frac{dp^0}{2\pi}
	\frac{\langle \alpha_1|\partial_{i_1} H |\alpha_2\rangle\langle \alpha_2 |\partial_{i_2} H | \alpha_3 \rangle
                \cdots \langle \alpha_{2n} |\partial_{i_{2n}} H | \alpha_1\rangle}
	{(ip^0+E_{\alpha_1})^2(ip^0+E_{\alpha_2})\cdots(ip^0+E_{\alpha_{2n}})},
\label{eq:J_2n+1}	
\end{eqnarray}
where $i_1,\cdots, i_{2n}$ stand for the spatial indices and summation over these indices are implicitly assumed following the Einstein's contraction rule. 
All we have to do is to integrate over $p^0$ using Cauchy's theorem. In order to discuss it in detail let us define the key integral $J$ as follows. 

\subsection{$p^0$ integration}
Here, we use a trick to simplify the integration. It is easy to see that the expression Eq.~(\ref{eq:winding_2n+1}) is invariant under continuous deformation of $S_F$ (or $H$) provided that the integrand remains to have no singularities. Therefore,  under  a continuous change of the Hamilitonian, the winding number remains unchanged from its original value as long as the enegry spectrum is kept gapped throughout the deformation. 

Now, the most general Hamiltonian with $N_v$ valence bands and $N_c$ conduction bands is expressed as
\begin{eqnarray}
H(\vec{p}) \equiv \sum_{a=1}^{N_v} E_a(\vec{p})  |a(\vec{p})\rangle \langle a(\vec{p}) |
+\sum_{\dot{b}=1}^{N_c} E_{\dot{b}}(\vec{p})  |\dot{b}(\vec{p})\rangle \langle \dot{b}(\vec{p}) |,
\end{eqnarray}
where $|a(\vec{p})\rangle$ labeled by $a$ is the energy eigenstate in the valence band with spatial momentum $\vec{p}$ and negative energy eigenvalue $E_a(\vec{p})<0$. 
The state $|\dot{b}(\vec{p})\rangle$ labeled by $\dot{b}$ is the energy eigenstate in the conduction band with spatial momentum $\vec{p}$ and positive energy eigenvalue $E_{\dot{b}}(\vec{p})>0 $. One can continuously deform the Hamiltonian without hitting the singularity of $S(p)$ ( i.e. keeping the system gapped ) so that all energy eigenvalues in the conduction bands and all energy eigenvalues in the valence bands are degenerate and momentum independent (i.e. flat band ) respectively.

Then the deformed Hamiltonian $H_{\rm new}$ which gives the same winding number becomes
\begin{eqnarray}
H_{\rm new}(\vec{p}) = 
E_v \sum_{a=1}^{N_v}  |a(\vec{p})\rangle \langle a(\vec{p}) |
+ E_c \sum_{\dot{b}=1}^{N_c}   |\dot{b}(\vec{p})\rangle \langle \dot{b}(\vec{p}) |,
\end{eqnarray}
where $E_v < 0$, $E_c>0$ are the momentum independent constant.  Here the eigenstates are identical to those with the original Hamiltonian.

Using the formulae Eqs. (\ref{eq:a_b}),  (\ref{eq:dota_dotb}) in Appendix ~\ref{useful_eigen}, one finds 
that in the insertion of eigenstates sandwitching $\partial_i H$, if states in the conduction bands appear in a row or if states in the valence bands appear in a row, 
the matrix element vanishes. Therefore, in Eq.(\ref{eq:J_2n+1}) states in the valence bands and the states in the conduction bands should appear in an alternating  order. Therefore, $J$ is expressed as
\begin{eqnarray}
J 
&=&
 \sum_{a_1,\cdots,a_n=1}^{N_v} \sum_{\dot{a}_1,\cdots,\dot{a}_n=1}^{N_c}  \epsilon^{i_1 j_1 \cdots i_{2n} j_{2n}} 
\nonumber\\
&&
\left[
\int \frac{dp^0}{2\pi} \frac{1}{(ip^0+E_v)^{n+1}(ip^0+E_c)^n}
	\langle a_1|\partial_{i_1} H | \dot{a}_1\rangle\langle \dot{a}_1 |\partial_{j_1} H | a_2\rangle \times
         \cdots       \times  \langle a_n |\partial_{i_n} H |\dot{a}_n\rangle\langle \dot{a}_n|\partial_{j_n} H | a_1\rangle
\right.
\nonumber\\
&&
+
\left.
\int \frac{dp^0}{2\pi} \frac{1}{(ip^0+E_c)^{n+1}(ip^0+E_v)^n}
	\langle \dot{a}_1|\partial_{i_1} H | a_1\rangle\langle a_1 |\partial_{j_1} H | \dot{a}_2\rangle \times
         \cdots       \times  \langle \dot{a}_n |\partial_{i_n} H |a_n\rangle\langle a_n|\partial_{j_n} H | \dot{a}_1\rangle
\right]
\nonumber\\
\end{eqnarray}
Renaming the labels for eigenstates and using the definition of master integral in Appendix~\ref{useful_p0} and substituting Eqs.(\ref{eq:a_dotb}),(\ref{eq:dota_b}), we have
\begin{eqnarray}
J 
&=&
 \sum_{a_1,\cdots,a_n=1}^{N_v} \sum_{\dot{a}_1,\cdots,\dot{a}_n=1}^{N_c}  \epsilon^{i_1 j_1 \cdots i_{2n} j_{2n}} 
(-1)^n (E_c-E_v)^{2n} \left( I^{[n+1,n]}(E_v,E_c)	- I^{[n+1,n]}(E_c,E_v)\right)
\nonumber\\
&&\langle a_1 | \partial_{i_1} \dot{a}_1\rangle \langle \dot{a}_1 | \partial_{j_1} a_2\rangle
\times \cdots  \times    \langle a_n|\partial_{i_n} \dot{a}_n\rangle\langle \dot{a}_n|\partial_{j_n} a_1\rangle ,
\label{eq:J_2n+1_abcd}	
\end{eqnarray}
where $I^{[m,n]}(E_1,E_2)$ is defined as
\begin{eqnarray}
I^{[m,n]}(E_1,E_2)
\equiv 
&
\displaystyle{ \int \frac{dp^0}{2\pi} 
\frac{1}{(ip^0+E_1)^m(ip^0+E_2)^n}  }.
\end{eqnarray}
The expression of $I^{[m,n]}(E_1,E_2)$ after $p^0$ integration is given in Appendix~\ref{useful_p0}.

Substituting Eq.~(\ref{eq:I_n+1_n}) into Eq.~(\ref{eq:J_2n+1_abcd}), we obtain
\begin{eqnarray}
J = - \sum_{a_1,\cdots,a_n=1}^{N_v} \sum_{\dot{a}_1,\cdots,\dot{a}_n=1}^{N_c}  \epsilon^{i_1 j_1 \cdots i_{2n} j_{2n}} 
 \frac{(2n)!}{(n!)^2}
\langle a_1 | \partial_{i_1} \dot{a}_1\rangle \langle \dot{a}_1 | \partial_{j_1} a_2\rangle
\times \cdots  \times    \langle a_n|\partial_{i_n} \dot{a}_n\rangle\langle \dot{a}_n|\partial_{j_n} a_1\rangle .
\nonumber\\
\end{eqnarray}
Using the formula for the Berry curvature in Eq.~(\ref{eq:Berry_curvature}) in Appendix~\ref{useful_eigen}, $J$ is finally expressed as
\begin{eqnarray}
J = (-1)^{n+1} \sum_{a_1,\cdots,a_n=1}^{N_v} \sum_{\dot{a}_1,\cdots,\dot{a}_n=1}^{N_c}  \epsilon^{i_1 j_1 \cdots i_{2n} j_{2n}} 
\frac{(2n)!}{(n!)^2} i^n \mathcal{F}^{a_1 a_2}_{i_1 j_1}\cdots\mathcal{F}^{a_n a_{n+1}}_{i_n j_n}
\label{eq:J_ch2n+1}
\end{eqnarray}
\subsection{Results of $c_{\rm cs}$}
Plugging Eq.(\ref{eq:J_ch2n+1}) into Eq.(\ref{eq:ccs_J2n+1}) 
\begin{eqnarray}
c_{\rm cs}
&=&  \frac{(-1)^n}{(n+1)!(2\pi)^n} \int_{BZ} \mbox{ch}_n(\mathcal{A}),
\end{eqnarray}
where $\mbox{ch}_n(\mathcal{A})$ is the 2nd Chern character defined by
\begin{eqnarray}
\mbox{ch}_n(\mathcal{A})&=&\frac{1}{n!}  \frac{1}{(2\pi)^n}
 \mbox{tr}( \mathcal{F}^n ). 
\end{eqnarray}
Comparing this expression with Eq.~(\ref{eq:ccs_k}) 
\begin{eqnarray}
c_{\rm cs} = \frac{k}{(n+1)!(2\pi)^n},
\end{eqnarray}
we arrive at the relation
\begin{eqnarray}
k= (-1)^n \int_{BZ}  \mbox{ch}_n(\mathcal{A}).
\end{eqnarray}
This means that the Chern-Simons level and the topological number in terms of the Berry connection is shown to be identical.

%
%
\section{Summary and Conclusion}
We derived general TKNN formula from Chern-Simons level in the effective action for 
lattice system with general Hamiltonian bilinear in fermion in (2+1)- and (4+1)- dimensions.
We have shown that Chern-Simons level is given by the winding number of a map from $T^D$ to the fermion propagator space. For this relation Ward-Takahashi identities including higher order relations are crucial.

There has been an understanding  that for the field theory approach to work, there should be a low energy mode which can be described by a relativistic field theory. Therefore, people had the impression that the field theory approach works only a special type of systems with emergent 
relativistic spectrum. 
The interesting point to note 
in our proof
is that one does not need to assume anything but gauge invariance. The detailed structure of minimal or non-minimal gauge coupling is irrelevant. 
Also one does not need to assume that there exists an effective theory described by the relativistic field theory 
and it applies to 
 any system including arbitrary bands which may be far away from the Fermi level. 

Since we have found that 
the two methods
can equally work well, 
 we are now certain that we can use field theory approach to study the topological properties for arbitrary condensed matter systems which include interactions where one can fully utilize the power of field theory. 

There are topological materials  
for systems with additional 
symmetries and in other dimensions 
Whether a complete equivalence holds for those systems remains an open problem. We hope to extend our study to those systems in the future.

\begin{acknowledgments}
We would like to thank Masatoshi Sato for his comments.  This work is supported in part by the Japanese Grant-in-Aid for Scientific Research(Nos. 15K05054, 18H01216, 18H04484 and 18K03620).
\end{acknowledgments}

%
%
%
\appendix

\section{Ward-Takahashi identities}
\label{App_WT}
In this appendix, we derive various identities among vertex functions and the inverse fermion propagator obtained from gauge invariance, i.e. Ward-Takahashi identities.
Finite difference operator which appears in the hopping term of the lattice fermion system can be expressed in terms of infinite series of derivatives. For example, consider a gauge invariant fermion bilinear term connected by a straight-line Wilson line in the $\mu$ direction as
\begin{eqnarray}
X\equiv \psi^\dagger(t,\vec{x}) e^{ i\int_{\vec{x}}^{\vec{x}+a\vec{\mu}} d\vec{r}^\prime\cdot\vec{A}(\vec{r}^\prime)} \psi(t,\vec{x}+a\vec{\mu}),
\end{eqnarray}
where $a$ is the lattice spacing and $\vec{\mu}$ is the unit vector in the $\mu$ direction.
This term can be formally expanded as
\begin{eqnarray}
X = a \sum_{n=0}^\infty
\psi^\dagger(t,\vec{x}) \sum_{n=0}^\infty \frac{a^n}{n!} \left(D^n_\mu  \psi\right)(t,\vec{x}),
\end{eqnarray}
where $D_\mu=\partial_\mu + i A_\mu$.
We assume that the Hamiltonian can be expressed in terms of all sorts of fermion hopping terms 
connected by the Wilson-lines of arbitrary contours or superpositions of them.
Then, the action can be formally expanded as
\begin{eqnarray}
S = \int dt \sum_{\vec{x}} \sum_{n=0}^\infty \psi^\dagger(t,\vec{x})
M_{\mu_1 \cdots \mu_n} (D_{\mu_1} \cdots D_{\mu_n}\psi)(t,\vec{x})
\label{eq:action}
\end{eqnarray}
where summation over $\mu_1, \cdots,\mu_n$ are implicit. $M_{\mu_1\cdots\mu_n}$ are some $N\times N$ matrix where $N=N_c+N_v$ is the number of fermion degrees of freedom per site.

Expanding this action in terms of gauge fields and making Fourier transformations, one can obtain the formal expressions of the inverse propagator and the vertex functions in the momentum space.
In the following, let us denote the inverse fermion propagator with momentum $p$ as $S^{-1}_F(p)$ and the vertex functions with incoming fermion momentum $p$ and $n$ photons with incoming momentum $k_i$ and $\mu_i$ components ($i=1,\cdots, n$) and outgoing fermion with momentum $p+ \displaystyle{\sum_{i=1}^n k_i}$ as $\Gamma^{(n)}[k_1,\mu_1;\cdots;k_n,\mu_n;p]$. Then the formal expression gives
\begin{eqnarray}
&&S_F^{-1}(p) = \sum_{n=0}^\infty M_{\mu_1\cdots\mu_n} \prod_{i=1}^n \left(ip_{\mu_i}\right),
\label{eq:S_F}
\end{eqnarray}
\begin{eqnarray}
&&\Gamma^{(1)}[k,\mu;p] =
-
i \sum_{n=1}^\infty \sum_{a=1}^nM_{\mu_1\cdots\mu_{a-1} \mu\mu_{a+1}\cdots \mu_n} \prod_{i=1}^{a-1} \left(i(p+k)_{\mu_i}\right) \prod_{i=a+1}^{n} \left(ip_{\mu_i}\right),
\label{eq:Gamma^{(1)}}
\end{eqnarray}
\begin{eqnarray}
&&\Gamma^{(2)}[k,\mu;l,\nu;p] 
\nonumber\\
&=& 
-
i^2\sum_{n=1}^\infty 
\sum_{\underset{a<b}{a,b=1}}^{n}
M_{\mu_1\cdots\mu_{a-1} \mu\mu_{a+1}\cdots \mu_{b-1}\nu\mu_{b+1}\cdots\mu_n} \prod_{i=1}^{a-1} \left(i(p+k+l)_{\mu_i}\right) \prod_{i=a+1}^{b-1} \left(i(p+l)_{\mu_i}\right) \prod_{i=b+1}^{n} \left(ip_{\mu_i}\right)
\nonumber\\
&
-
&i^2\sum_{n=1}^\infty 
\sum_{\underset{a<b}{a,b=1}}^{n}
M_{\mu_1\cdots\mu_{a-1} \nu\mu_{a+1}\cdots \mu_{b-1}\mu\mu_{b+1}\cdots\mu_n} \prod_{i=1}^{a-1} \left(i(p+k+l)_{\mu_i}\right) \prod_{i=a+1}^{b-1} \left(i(p+k)_{\mu_i}\right) \prod_{i=b+1}^{n} \left(ip_{\mu_i}\right),
\nonumber\\
\label{eq:Gamma^{(2)}}
\end{eqnarray}
\begin{eqnarray}
&&\Gamma^{(3)}[k,\mu;l,\nu, r,\lambda;p] 
\nonumber\\
&=& 
-
i^3\sum_{n=1}^\infty 
\sum_{\underset{a<b<c}{a,b,c=1}}^{n}
M_{\mu_1\cdots\mu_{a-1} ~\mu ~\mu_{a+1}\cdots \mu_{b-1}~\nu~\mu_{b+1}\cdots\mu_{c-1}
~\lambda~\mu_{c+1}\cdots \mu_{n}
}
\nonumber\\
&&\prod_{i=1}^{a-1} \left(i(p+k+l+r)_{\mu_i}\right) \prod_{i=a+1}^{b-1} \left(i(p+l+r)_{\mu_i}\right) \prod_{i=b+1}^{c-1} \left(i(p+r)_{\mu_i}\right)
\prod_{i=c+1}^{n} \left(ip+r_{\mu_i}\right)
\nonumber\\
&&
+[ \mbox{ other 5 terms obtained from the  permutaion of }
(\mu, k), (\nu,l),(\lambda,r) ] .
\label{eq:Gamma^{(3)}}
\end{eqnarray}
Differentiating  Eq.(\ref{eq:S_F}) with respect to $p_\mu$ and taking soft photon limit ($k\rightarrow 0$) in Eq. (\ref{eq:Gamma^{(1)}}), one obtains
\begin{eqnarray}
\frac{\partial S^{-1}_F(p)}{\partial p_\mu} 
= 
-
\left. \Gamma^{(1)}[k,\mu;p]\right|_{k=0}
=i \sum_{n=1}^\infty \sum_{a=1}^nM_{\mu_1\cdots\mu_{a-1} \mu\mu_{a+1}\cdots \mu_n} \prod_{i\neq a}^{1\sim n} \left(ip_{\mu_i}\right).
\end{eqnarray}
This is the well-known Ward-Takahashi identity in QED, which is generalized for the lattice fermion system.

\subsection{First order Ward-Takahashi identities}
It is interesting to note that we could also obtain Ward-Takahashi identities for quantities involving higher order terms in photon momenta and multi-photon vertex functions. In order to see that, let us take the second derivatives of Eq. (\ref{eq:S_F}). One obtains
\begin{eqnarray}
\frac{\partial^2 S^{-1}_F(p)}{\partial p_\mu \partial p_\nu}
&=&
i^2\sum_{n=1}^\infty 
\sum_{\underset{a<b}{a,b=1}}^{n}
\left[M_{\mu_1\cdots\mu_{a-1} \mu\mu_{a+1}\cdots \mu_{b-1}\nu\mu_{b+1}\cdots\mu_n} 
+(\mu\leftrightarrow\nu) \right]
\prod_{i\neq a,b}^{1\sim n}  \left(ip_{\mu_i}\right)
\end{eqnarray}
Let us also differentiate Eq.(\ref{eq:Gamma^{(1)}}) with $k_\nu$ or $p_\nu$ and take the soft photon limit. 
We obtain
\begin{eqnarray}
\left. \frac{\partial \Gamma^{(1)}[k,\mu;p]}{\partial k_\nu}\right|_{k=0}
&=&
-
i^2\sum_{n=1}^\infty 
\sum_{\underset{a<b}{a,b=1}}^{n}
M_{\mu_1\cdots\mu_{a-1} \nu\mu_{a+1}\cdots \mu_{b-1}\mu\mu_{b+1}\cdots\mu_n} 
\prod_{i\neq a,b}^{1\sim n}  \left(ip_{\mu_i}\right)
\\
\left.\frac{\partial \Gamma^{(1)}[k,\mu;p]}{\partial p_\nu}\right|_{k=0}
&=&
-
i^2\sum_{n=1}^\infty 
\sum_{\underset{a<b}{a,b=1}}^{n}
\left[M_{\mu_1\cdots\mu_{a-1} \mu\mu_{a+1}\cdots \mu_{b-1}\nu\mu_{b+1}\cdots\mu_n} 
+(\mu\leftrightarrow\nu) \right]
\prod_{i\neq a,b}^{1\sim n}  \left(ip_{\mu_i}\right)
\end{eqnarray}
Taking also the soft photon limit of Eq.(\ref{eq:Gamma^{(2)}}), we obtain
\begin{eqnarray}
\left.\Gamma^{(2)}[k,\mu;l,\nu;p]\right|_{k,l=0}
&=&
-
i^2\sum_{n=1}^\infty 
\sum_{\underset{a<b}{a,b=1}}^{n}
\left[M_{\mu_1\cdots\mu_{a-1} \mu\mu_{a+1}\cdots \mu_{b-1}\nu\mu_{b+1}\cdots\mu_n} 
+(\mu\leftrightarrow\nu)\right]
\prod_{i\neq a,b}^{1\sim n}  \left(ip_{\mu_i}\right)
\nonumber\\
\end{eqnarray}
We then obtain the following identities:
\begin{eqnarray}
\frac{\partial^2 S^{-1}_F(p)}{\partial p_\mu \partial p_\nu}
=
-
\left. \left(\frac{\partial \Gamma^{(1)}[k,\mu;p]}{\partial k_\nu}
+(\mu\leftrightarrow\nu)\right)\right|_{k=0}
=
-
\left. \frac{\partial \Gamma^{(1)}[k,\mu;p]}{\partial p_\nu}\right|_{k=0}
=
-
\left.\Gamma^{(2)}[k,\mu;l,\nu;p]\right|_{k,l=0}
\nonumber\\
\end{eqnarray}
\subsection{Second order Ward-Takahashi identities}
We could go even further in higher order. Taking the third derivative of Eq.(\ref{eq:S_F}),  we obtain
\begin{eqnarray}
\frac{\partial^3 S^{-1}_F(p)}{\partial p_\mu \partial p_\nu \partial p_\lambda}
&=&
i^3\sum_{n=1}^\infty 
\sum_{\underset{a<b<c}{a,b,c=1}}^{n}
\left[M_{\mu_1\cdots\mu_{a-1} \mu\mu_{a+1}\cdots \mu_{b-1}\nu\mu_{b+1}\cdots \mu_{c-1}\lambda\mu_{c+1}\cdots \mu_n} 
+(\mbox{perm. in}(\mu,\nu,\lambda)) \right]
\nonumber\\
&&
\times\prod_{i\neq a,b,c}^{1\sim n}  \left(ip_{\mu_i}\right)
\end{eqnarray}
The second derivatives of Eq.(\ref{eq:Gamma^{(1)}}) with respect to $k$ or $p$ give
\begin{eqnarray}
\left. \frac{\partial^2 \Gamma^{(1)}[k,\mu;p]}{\partial k_\nu \partial k_\lambda}\right|_{k=0}
&=&
-
i^3\sum_{n=1}^\infty 
\sum_{\underset{a<b<c}{a,b,c=1}}^{n}
\left[M_{\mu_1\cdots\mu_{a-1} \nu\mu_{a+1}\cdots \mu_{b-1}\lambda\mu_{b+1}\cdots \mu_{c-1}\mu\mu_{c+1}\cdots \mu_n} 
+
(\nu\leftrightarrow\lambda)
\right]
\nonumber\\
&&\times\prod_{i\neq a,b,c}^{1\sim n}  \left(ip_{\mu_i}\right)
\\
\left. \frac{\partial^2 \Gamma^{(1)}[k,\mu;p]}{\partial k_\nu \partial p_\lambda}\right|_{k=0}
&=&
-
i^3\sum_{n=1}^\infty 
\sum_{\underset{a<b<c}{a,b,c=1}}^{n}
\left[M_{\mu_1\cdots\mu_{a-1} \nu\mu_{a+1}\cdots \mu_{b-1}\lambda\mu_{b+1}\cdots \mu_{c-1}\mu\mu_{c+1}\cdots \mu_n} 
+(\nu\leftrightarrow\lambda) \right.
\nonumber\\
&&\left.
+((\nu,\lambda,\mu)
\leftrightarrow
(\nu,\mu,\lambda)) 
 \right]
 \prod_{i\neq a,b,c}^{1\sim n}  \left(ip_{\mu_i}\right)
\\
\left. \frac{\partial^2 \Gamma^{(1)}[k,\mu;p]}{\partial p_\nu \partial p_\lambda}\right|_{k=0}
&=&
-
i^3\sum_{n=1}^\infty 
\sum_{\underset{a<b<c}{a,b,c=1}}^{n}
\left[M_{\mu_1\cdots\mu_{a-1} \mu\mu_{a+1}\cdots \mu_{b-1}\nu\mu_{b+1}\cdots \mu_{c-1}\lambda\mu_{c+1}\cdots \mu_n} 
+(\mbox{perm. in}(\mu,\nu,\lambda)) \right]
\nonumber\\
&& \times\prod_{i\neq a,b,c}^{1\sim n}  \left(ip_{\mu_i}\right)
\end{eqnarray}
From the first derivative of Eq.(\ref{eq:Gamma^{(2)}}) with respect to $k$ or $l$ or $p$, we obtain
\begin{eqnarray}
\left. \frac{\partial \Gamma^{(2)}[k,\mu; l,\nu;p]}{\partial k_\lambda}\right|_{k,l=0}
&=&
-
i^3\sum_{n=1}^\infty 
\sum_{\underset{a<b<c}{a,b,c=1}}^{n}
\left[M_{\mu_1\cdots\mu_{a-1} \lambda\mu_{a+1}\cdots \mu_{b-1}\mu\mu_{b+1}\cdots \mu_{c-1}\nu\mu_{c+1}\cdots \mu_n} \right.
\nonumber\\
&&
\left.
+((\lambda,\mu,\nu)
\leftrightarrow(
\lambda,\nu,\mu))
+((\lambda,\mu,\nu)
\leftrightarrow
(\nu,\lambda,\mu))
 \right]
\prod_{i\neq a,b,c}^{1\sim n}  \left(ip_{\mu_i}\right)
\nonumber\\
\\
\left. \frac{\partial \Gamma^{(2)}[k,\mu; l,\nu;p]}{\partial l_\lambda}\right|_{k,l=0}
&=&
-
i^3\sum_{n=1}^\infty 
\sum_{\underset{a<b<c}{a,b,c=1}}^{n}
\left[M_{\mu_1\cdots\mu_{a-1} \lambda\mu_{a+1}\cdots \mu_{b-1}\mu\mu_{b+1}\cdots \mu_{c-1}\nu\mu_{c+1}\cdots \mu_n} \right.
\nonumber\\
&&
\left.
+((\lambda,\mu,\nu)
\leftrightarrow
(\mu,\lambda,\nu))
+((\lambda,\mu,\nu)
\leftrightarrow
(\lambda,\nu,\mu))
 \right]
\prod_{i\neq a,b,c}^{1\sim n}  \left(ip_{\mu_i}\right)
\nonumber\\
\\
\left. \frac{\partial \Gamma^{(2)}[k,\mu;l,\nu;p]}{\partial p_\lambda}\right|_{k,l=0}
&=&
-
i^3\sum_{n=1}^\infty 
\sum_{\underset{a<b<c}{a,b,c=1}}^{n}
\left[M_{\mu_1\cdots\mu_{a-1} \mu\mu_{a+1}\cdots \mu_{b-1}\nu\mu_{b+1}\cdots \mu_{c-1}\lambda\mu_{c+1}\cdots \mu_n} \right.
\nonumber\\
&&+\left. (\mbox{perm. in}(\mu,\nu,\lambda)) \right]
 \prod_{i\neq a,b,c}^{1\sim n}  \left(ip_{\mu_i}\right)
\end{eqnarray}
These equations give the following identities:
\begin{eqnarray}
\left.\frac{\partial^2 \Gamma^{(1)}[k,\mu;p]}{\partial k_\nu \partial p_\lambda}\right|_{k=0}
=\left. \frac{\partial \Gamma^{(2)}[k,\mu;l,\lambda;p]}{\partial k_\nu}\right|_{k,l=0}
=\left. \frac{\partial \Gamma^{(2)}[k,\lambda;l,\mu;p]}{\partial l_\nu}\right|_{k,l=0}
\end{eqnarray}
\subsection{Third order Ward-Takahashi identities}
Carrying out similar calculations by simply differentiaing $\Gamma^{(2)}$ and $\Gamma^{(3)}$ 
given in Eq.(\ref{eq:Gamma^{(2)}}) and Eq.(\ref{eq:Gamma^{(3)}}), 
we can see that the following identity holds:
\begin{eqnarray}
\left.\frac{\partial^2 \Gamma^{(3)}[q,\mu; r,\nu;s,\lambda;p]}{\partial q_{\alpha} \partial r_{\beta}}\right|_{q,r,s=0}
= \left.\frac{\partial^3\Gamma^{(2)}[q,\mu; r,\nu;p]}{\partial q_{\alpha} \partial r_{\beta} \partial p_\lambda}\right|_{q,r=0}
\end{eqnarray}

\section{Useful formulae for energy eigenstate}\label{useful_eigen}
Consider a Hamiltonian in momentum representation $H(p)$ and the normalized complete set of eigenstates at momentum $p$ labeled by index $\alpha$ 
($\alpha$ runs over the states in both the valence and the conduction bands) with the following properties:
\begin{eqnarray}
H(p) | \alpha(p)\rangle &=& E_\alpha(p) | \alpha(p)\rangle, 
\label{eq:eigen}\\
\langle \alpha | \beta \rangle &=& \delta_{\alpha\beta}, 
\label{eq:orthonormal}\\
\langle \alpha(p) | H(p) | \beta(p)\rangle &=& E_\alpha(p) \delta_{\alpha\beta}, ~~~~~~~(\alpha,\beta=1,\cdots, N).
\label{eq:eigenmatrix}
\end{eqnarray}
Let us consider  differentiation with respect to $p_\mu$.  Here we introduce the simplified notation
\begin{eqnarray}
|\partial_\mu \alpha\rangle = \partial_\mu | \alpha \rangle \equiv \frac{\partial}{\partial p_\mu} \left(| \alpha(p)\rangle\right),
&&
\partial_\mu H \equiv  \frac{\partial}{\partial p_\mu}\left( H(p)\right).
\end{eqnarray}
Then differentiating Eq.(\ref{eq:orthonormal}), we obtain
\begin{eqnarray}
\langle \partial_\mu \alpha | \beta \rangle = - \langle \alpha | \partial_\mu \beta\rangle .
\label{eq:ab_deriv}
\end{eqnarray}
Also, differentiating Eq.(\ref{eq:eigenmatrix}) and making a little algebra, we have
\begin{eqnarray}
(E_\alpha - E_\beta ) \langle \alpha | \partial_\mu \beta \rangle + \langle \alpha | \partial_\mu H | \beta\rangle = \partial_\mu E_\alpha \delta_{\alpha\beta}.
\end{eqnarray}
This means the matrix element of the momentum derivative of the Hamiltonian is given as
\begin{eqnarray}
 \langle \alpha| \partial_\mu H | \alpha \rangle &=& \partial_\mu E_\alpha,
\label{eq: H_deriv_aa}
 \\
  \langle \alpha| \partial_\mu H | \beta \rangle &=& -(E_\alpha - E_\beta) \langle \alpha | \partial_\mu \beta \rangle ~~~~~~(\alpha\ne \beta).
\label{eq: H_deriv_ab}
\end{eqnarray}

\subsection{Degenerate flat band case}
Let us now consider the special case where all the energies in the valence bands and those in the conduction bands are degenerate  and flat.
Then, one can easily see that 
\begin{eqnarray}
\langle a (\vec{p}) | \partial_\mu H(\vec{p}) | b(\vec{p})\rangle  &=& 0
\label{eq:a_b}
\\
\langle \dot{a} (\vec{p}) | \partial_\mu H(\vec{p}) | \dot{b}(\vec{p})\rangle  &=& 0 ,
\label{eq:dota_dotb}
\\
\langle a (\vec{p}) | \partial_\mu H(\vec{p}) | \dot{b}(\vec{p})\rangle  &=& (E_c - E_v) \langle a | \partial_\mu \dot{b}\rangle, 
\label{eq:a_dotb}
\\
\langle \dot{a} (\vec{p}) | \partial_\mu H(\vec{p}) | b(\vec{p})\rangle  &=& -(E_c - E_v) \langle \dot{a} | \partial_\mu  b\rangle, ~~~(a,b=1,\cdots, N_v,  \dot{a},\dot{b}=1,\cdots,N_c).
\label{eq:dota_b}
\end{eqnarray}
where the states with undotted indices  $|a\rangle$, $|b\rangle$ are  in the valance bands with constant energy $E_v<0$  and 
those with dotted indices $|\dot{a}\rangle$, $|\dot{b}\rangle$ are  in the conduction bands  with constant energy $E_c>0$.

Let us define the Berry connection using the negative energy eigenstates as
\begin{eqnarray}
\mathcal{A}^{ab} \equiv \mathcal{A}^{ab}_\mu dx^\mu = - i \langle a | \partial_\mu b \rangle dx^\mu
\equiv -i \langle a | db\rangle .
\end{eqnarray}
Then the Berry curvature $\mathcal{F}^{ab}$ is 
\begin{eqnarray}
\mathcal{F}^{ab} &=&\left( d\mathcal{A} + i \mathcal{A} \mathcal{A} \right)^{ab}
= -i \langle d a| d b\rangle +i \sum_{c=1}^{N_v}(-i) \langle a | dc\rangle(-i) \langle c | db\rangle
\nonumber\\
&=& -i \left[\sum_{c=1}^{N_v} \langle d a| c\rangle \langle c | d b\rangle  
      + \sum_{\dot{c}=1}^{N_c} \langle d a| \dot{c}\rangle \langle \dot{c} | d b\rangle  \right]
-  i\sum_{c=1}^{N_v} \langle a | dc\rangle \langle c | db\rangle
\nonumber\\
&=& i \sum_{c=1}^{N_v} \langle  a| d c\rangle \langle c | d b\rangle  
      +i \sum_{\dot{c}=1}^{N_c} \langle a| d\dot{c}\rangle \langle \dot{c} | d b\rangle  )
-  i\sum_{c=1}^{N_v} \langle a | dc\rangle \langle c | db\rangle
\nonumber\\
&=&       i \sum_{\dot{c}=1}^{N_c} \langle a| d\dot{c}\rangle \langle \dot{c} | d b\rangle 
\label{eq:Berry_curvature}
\end{eqnarray}


\section{Useful formulae for $p^0$ integration}\label{useful_p0}
We define the following $p^0$ integrations:
\begin{eqnarray}
I^{[m,n]}(E_1,E_2)
\equiv 
&
\displaystyle{ \int \frac{dp^0}{2\pi} 
\frac{1}{(ip^0+E_1)^m(ip^0+E_2)^n}  }
~~& (E_1\ne E_2)
\end{eqnarray}
Simple contour integral by adding a contour in the lower half semi-circle in the complex $p^0$ plane and picking up poles in the lower half plane,
we obtain
\begin{eqnarray}
I^{[m,n]}(E_1,E_2) &=& (-1)^m \frac{(m+n-2)!}{(m-1)!(n-1)!} \cdot\frac{\theta(-E_1)-\theta(-E_2)}{(-E_1+E_2)^{m+n-1}}.
\end{eqnarray} 
In particular, when $(m,n)=(n+1,n)$ 
\begin{eqnarray}
I^{[n+1,n]}(E_1,E_2) &=& (-1)^{n+1} \frac{(2n)!}{2(n!)^2} \cdot\frac{\theta(-E_1)-\theta(-E_2)}{(-E_1+E_2)^{2n}}
\label{eq:I_n+1_n}
\end{eqnarray} 
holds. This formula is useful in making $p^0$ integration of the propagator expressions for Chern-Simons level.


\bibliography{Ref}
\bibliographystyle{junsrt}

\end{document}